\documentclass[structabstract]{aa}
\usepackage{graphicx}
\usepackage{txfonts}
\usepackage{float}
\usepackage{natbib}
\usepackage[english]{babel}
\bibpunct{(}{)}{;}{a}{}{,}
\begin{document}

\title{CXOM31~J004253.1+411422: The first ultra$-$luminous X$-$ray transient in M~31\thanks{Based on observations obtained with 
   XMM-Newton, an ESA Science Mission with instruments and contributions 
   directly funded by ESA Member States and NASA.}}

\author{A. Kaur\inst{1}, M. Henze\inst{2}, F. Haberl\inst{2}, W. Pietsch\inst{2}, J. Greiner\inst{2}, A. Rau\inst{2}, D.H. Hartmann\inst{1}, G. Sala\inst{3}, M. Hernanz\inst{4}}

%\author{A. Kaur\inst{1}, M. Henze\inst{2}, W. Pietsch\inst{2}, V. Burwitz\inst{2}, M. Della Valle\inst{3,}\inst{4,}\inst{5}, M. Freyberg\inst{2}, J. Greiner\inst{2}, M. Hernanz\inst{6}, F. Haberl\inst{2}, D.H. Hartmann\inst{1}, D. Hatzidimitriou\inst{7,}\inst{8}, A. Riffeser\inst{10}, G. Sala\inst{9}, S. Seitz\inst{2,}\inst{10}, P. Reig\inst{8,}\inst{11}, A. Rau\inst{2}}

\institute{Department of Physics and Astronomy, Clemson University, Clemson, SC 29634 
\and Max$-$Planck$-$Institut f{\"u}r extraterrestrische Physik, Giessenbachstrasse, 85748 Garching, Germany
\and Department of F\'{i}sica i Enginyeria Nuclear, EUETIB (UPC-IEEC), Comte d'Urgell 187, 08036 Barcelona, Spain
%\and European Southern Observatory (ESO), D-85748 Garching, Germany
%\and INAF-Napoli, Osservatorio Astronomico di Capodimonte, Salita Moiariello 16 I-80131 Napoli, Italy
%\and International Centre for Relativistic Astrophysics, Piazzale della Repubblica 2, I-65122 Pescara, Italy
\and Institut de Ci\'{e}ncies de 1'Espai (CSIC-IEEC), Campus UAB, Facultat Ci\'{e}ncies, C5 parell 2$^{on}$, 08193 Bellaterra (Barcelona), Spain}
%\and Department of Astrophysics, Astronomy and Mechanics, Faculty of Physics, University of Athens, Panepistimiopolis, GR 15784 Zografos, Athens, Greece 
%\and Foundation of Research and Technology Hellas, IESL, Greece
%\and Department of F\'{i}sica i Enginyeria Nuclear, EUETIB (UPC-IEEC), Comte d'Urgell 187, 08036 Barcelona, Spain 
%\and University Observatory Munich, Scheinerstrasse 1, 81679 Munich, Germany
%\and Department of Physics and Institute of theoratical \& Computational Physics, University of Crete, GR-71003 Heraklion, Greece }

\date{}

\abstract 
{We seek clarification of the nature of X$-$ray sources detected in M~31. Here we focus on CXOM31~J004253.1+411422, the brightness of which suggests that it belongs to the class of ultraluminous X$-$ray sources.}
{We determine the X$-$ray properties of sources detected in the XMM$-$Newton / $\it{Chandra}$ monitoring program. We investigate spectral properties and search for periodic or quasi$-$periodic oscillations. A multi$-$component model is applied to the spectra obtained from XMM$-$Newton data to evaluate the relative contributions from thermal and non$-$thermal emission. The time dependence of this ratio is evaluated over a period of forty days.}
{We simultaneously fit data from XMM$-$Newton EPIC-pn, MOS1 and MOS2 detectors with (non$-$thermal) powerlaw  and  (thermal) multicolored blackbody.}
{The X$-$ray spectrum is best fit by the combination of a thermal component with kT $\sim$ 1 keV and a powerlaw component with photon index approximately 2.6. From combined analysis of $\it{Chandra}$, $\it{Swift}$ and XMM$-$Newton data, the unabsorbed total luminosity of this source decreases from $\sim$ 3.8 x 10$^{39}$ erg s$^{-1}$ in the first observation to $\sim$ 0.5 x 10$^{39}$ ergs s$^{-1}$ over a period of three months. The decay closely follows an exponential decline with a time constant of 32 days. The source spectrum evolves significantly, exhibiting a faster decline of the thermal component. We do not find evidence of any significant temporal features in the power density spectrum. The presence of a thermal component at kT $\sim$ 1 keV in conjunction with a non$-$thermal high energy tail, is also consistent with spectral properties of other ULXs in the "high state".}
{Our analysis indicates that the underlying source of this first ULX in M~31 is a black hole of mass, M $\geq$ 13 M$_{\sun}$, accreting near the Eddington limit, that underwent a transient outburst followed by an exponential decay reminiscent of transients associated with galactic X$-$ray novae.}

\keywords{Galaxies: individual: M~31 $-$ X$-$rays: stars $-$ X$-$rays: binaries}
\titlerunning{CXOM31~J004253.1+411422:The first ULX in M~31}
\authorrunning{A. Kaur et al.}
\maketitle 
\section{Introduction}

Ultraluminous X$-$ray sources (ULXs) are very bright point sources with an X$-$ray luminosity of L$_{x} > 10^{39}$erg s$^{-1}$, which exceeds the Eddington luminosity  for compact objects with mass of approximately 10 M$_{\sun}$. This source class was first identified by the \emph{Einstein Observatory} in the eighties \citep{1989ARA&A..27...87F}. These sources are not associated with the center of galaxies and thus do not belong to the class of super$-$massive black holes. On the other hand, they are too bright to be associated with stellar mass black holes with M $<$ 10 M$_{\sun}$, accreting below the Eddington rate. An exciting possibility is that the underlying sources of ULXs are intermediate mass black holes (IMBHs), in the mass range $10^{2}$ $-$ $10^{4}$$M_{\sun}$, accreting at sub$-$Eddington rates \citep{2004IJMPD..13....1M}. For example, observations of an ULX in NGC5408 suggested a mass of $\sim$ 100 M$_{\sun}$, indicating a possible existence of IMBHs \citep{1993MNRAS.263L..51F}. However, a recent study of this source's variability indicates that the mass is likely below 100 M$_{\sun}$ \citep{2011AN....332..388M}. Although the identification of ULXs as IMBHs has not yet been firmly established, one should contemplate the theoretical progenitors, that could result in producing such objects. Several theories have been suggested for the formation of IMBH. \citet{2001ApJ...551L..27M} studied the collapse of Population III stars, while others considered the collapse of massive stars (in young super$-$massive star clusters) e.g. \citet{2004MNRAS.355..413P}. Another possibility is of course that the masses are much less than 100 M$_{\sun}$, but that the source accretes  at Super$-$Eddington rate, as suggested by \citet{2002ApJ...568L..97B}. Yet another factor to be considered is deviations from the common assumption of isotropic emission as suggested by \citet{2001ApJ...552L.109K}. Along these lines for anisotropic emission, \citet{1997MNRAS.286..349R} proposed a link between ULXs and extreme beaming, as observed in galactic micro$-$quasars. The nature of the ULX class remains unclear, which is in part due to the fact that the sample size is still rather small.

To better understand the ULX phenomenon, an increase in sample size and careful consideration of spectral and sample properties is desirable. The ULX sources  are mainly found in star forming galaxies. Until recently, no ULX had been identified in M~31. Our ongoing monitoring program$\footnote{www.mpe.mpg.de/m31novae/xray/index.php}$ for resolving super soft source states of optical novae in the central area of M~31  (P.I. Wolfgang Pietsch) with XMM$-$Newton and $\it{Chandra}$~HRC-I yielded a transient source, which, as discussed in this paper, represents the first recognized ULX in our companion galaxy.

We present the time development of the outburst and the spectral analysis of a ULX in M~31 using XMM$-$Newton, $\it{Chandra}$ and $\it{Swift}$ data over a period of approximately three months (Sect. 2), followed by the summary of the results in Sect. 3. Conclusions are presented in Sect. 4.
%More details about this proposal can be found at the link given below$$www.mpe.mpg.de/~m31novae/xray/index.php$$\textbf{talk about light curve from Barnard's paper and CXOM31 detection in X-ray and optical.. Henze et al etc..}%

\begin{table*}[ht!]
 \caption[]{Observations log}
\label{Table:1}
\centering 
         \begin{tabular}{llllll}
            \hline
\hline
          \\
Telescope/instrument & Obs ID & $~$Date & Exptime & Rate$^{a}$ & L$_{unabs}^{g}$\\

& & (UT) & (ks) & (ct s$^{-1}$) & (10$^{39}$ erg s$^{-1}$)   \\
	  \hline
\hline
\\
$\it{Chandra}$~HRC-I & 10885 & 2009$-$12$-$08.94 & 18.27 & $<$ 1.5e-03 & 0.002$^{e}$ \\
$\it{Chandra}$~HRC-I&10886 & 2009$-$12$-$17.90  & 18.34 & 3.300 $~\pm~$ 0.040 & 3.77 $~\pm~$0.04\\
$\it{Swift}-$XRT & 00031518013 & 2009$-$12$-$22.04 & 3.6 &  0.660 $~\pm~$ 0.200 & 3.04 $~\pm~$0.20 \\
$\it{Swift}-$XRT& 00035336016 & 2009$-$12$-$23.05 & 4.2 & 0.720 $~\pm~$ 0.100 & 3.31 $~\pm~$0.10 \\
$\it{Swift}-$XRT& 00035336017 & 2009$-$12$-$24.04 & 4.8 & 0.620 $~\pm~$ 0.100 & 2.86 $~\pm~$0.10 \\
$\it{Swift}-$XRT& 00035336018 & 2009$-$12$-$25.05 & 5.2 & 0.600 $~\pm~$ 0.100 & 2.76 $~\pm~$0.10 \\
$\it{Swift}-$XRT& 00035336019 & 2009$-$12$-$26.26 & 5.0 & 0.560 $~\pm~$ 0.100 & 2.58 $~\pm~$0.10 \\
$\it{Swift}-$XRT& 00035336020 & 2009$-$12$-$27.07 & 5.0 & 0.580 $~\pm~$ 0.100 & 2.67 $~\pm~$0.10 \\
XMM$-$Newton & 0600660201 & 2009$-$12$-$28.53  & 16.86  & 6.447 $~\pm~$ 0.022$^{b}$ & 2.16 $~\pm~$0.07$^{f}$ \\
&&&& 1.810 $~\pm~$ 0.010$^{c}$   \\
&&&& 1.851 $~\pm~$ 0.010$^{d}$  \\
XMM$-$Newton & 0600660301 & 2010$-$01$-$07.32 & 15.43 & 1.784 $~\pm~$ 0.012 & 1.49 $~\pm~$0.05$^{f}$ \\
&&&  & 1.105 $~\pm~$ 0.008  \\
&&&& 1.143 $~\pm~$ 0.008  \\
XMM$-$Newton & 0600660401 & 2010$-$01$-$15.53 & 15.33 & 3.832 $~\pm~$ 0.017 & 1.16 $~\pm~$0.04$^{f}$ \\
&&& & 1.372 $~\pm~$ 0.009 \\
&&& & 1.402 $~\pm~$ 0.009 \\
XMM$-$Newton & 0600660501 & 2010$-$01$-$25.11 & 17.83 & 3.042 $~\pm~$ 0.015 & 0.71 $~\pm~$0.03$^{f}$\\
&&&  & 0.846 $~\pm~$ 0.007  \\
&&&  & 0.873 $~\pm~$ 0.007 \\
XMM$-$Newton & 0600660601 & 2010$-$02$-$02.11 & 15.43 & 1.072 $~\pm~$ 0.009 & 0.65 $~\pm~$0.03$^{f}$ \\
&&& & 0.694 $~\pm~$ 0.006\\
&&&  & 0.687 $~\pm~$ 0.006  \\
$\it{Chandra}$~HRC-I&10808 & 2010$-$02$-$15.86 & 17.12 & 0.547 $~\pm~$ 0.080 & 0.51 $~\pm~$0.08 \\
$\it{Chandra}$~HRC-I&11809 & 2010$-$02$-$26.27 & 18.42 & 0.490 $~\pm~$ 0.180 & 0.46 $~\pm~$0.18 \\
\hline
\hline
\end{tabular}
\tablefoot{
\tablefoottext{a}{Count rate as observed from the source without rejecting pile-up pixels.}
\tablefoottext{b}{EPIC-pn,}
\tablefoottext{c}{MOS1,}
\tablefoottext{d}{MOS2 data.}  
\tablefoottext{e}{3$\sigma$ upper limit}
\tablefoottext{f}{Derived from the simultaneous fitting of EPIC$-$pn, MOS1 and MOS2 in XSPEC.}    
\tablefoottext{g}{Total unabsorbed luminosity from the source within (0.2$-$10 keV) energy band.}
 }
  \end{table*}

\begin{figure}[H]  
\centering
  \includegraphics[angle=0,scale=0.45]{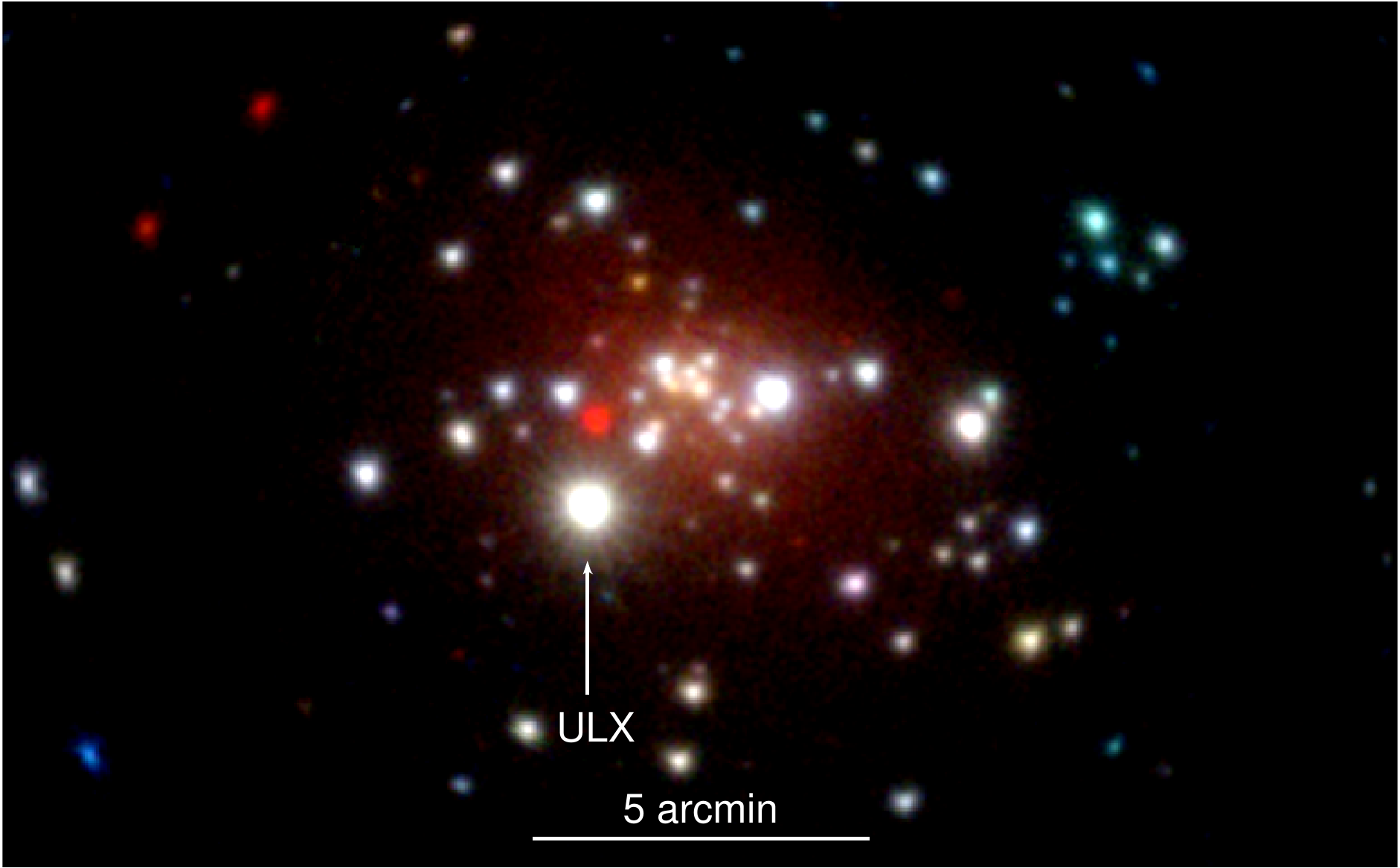}     

\caption{XMM-Newton EPIC image of the central part of M31 produced by combining pn, MOS1 and MOS2 data from all 5 observations. Red, green and blue show the (0.2 $-$ 1.0) keV, (1.0 $-$ 2.0) keV) and (2.0 $-$ 12.0) keV energy bands, respectively.}
               
 \end{figure}
\section{Observations and data analysis}
CXOM31~J004253.1+411422 was discovered with $\it{Chandra}$~HRC$-$I in a 19 ks observation on 2009 Dec 17.89 UT \citep{2009ATel.2356....1H}. The source position was determined to be R.A (J2000) = 00:42:53.15, Dec. (J2000) = +41:14:22.9, using the catalog of X$-$ray sources in M~31 assembled by \citet{2002ApJ...578..114K} for relative astrometry. 
Figure 1 shows the position of the source with respect to other X$-$ray sources near the center of M31 in an XMM$-$Newton color image. A transient optical counterpart with m(F435W) = 23.8 mag was identified in HST observations \citep{2010ATel.2474....1G}. The X$-$ray source was first detected with $\it{Chandra}$~HRC-I. Level 2 event files were analyzed using \citep[Chandra Interactive Analysis of 
    Observations;][]{2006SPIE.6270E..60F}$\footnote{http://cxc.harvard.edu/ciao/}$ to obtain count rates as shown in Table 1.  An adapted version of the XMMSAS tool \texttt{emldetect} was used to estimate background-corrected and exposure-corrected fluxes and count rates \citep[see][]{2010A&A...523A..89H}. Continuous monitoring of this source was then carried out by $\it{Swift}-$XRT from December 22 $-$ 27, 2009. The data were analysed using the HEAsoft XIMAGE package (version 4.5.1) with the \texttt{sosta} command (source statistics) for estimations of count rates. We took into account the XRT PSF of the sources that we computed with the command \texttt{psf}, as well as exposure maps that were created with the XRT software task \texttt{xrtexpomap} within XIMAGE. The count rates obtained from $\it{Swift}-$XRT and $\it{Chandra}$~HRC-I before and after the XMM$-$Newton observations were converted to unabsorbed fluxes using energy conversion factors (ecfs), which were computed using \texttt{fakeit} in XSPEC assuming the best fitting spectral model for the first and last XMM$-$Newton observations, respectively, and using publicly available instrument response files. We obtain ecf$_{HRC-I-1}$ = 6.3 x 10$^{10}$ cts erg$^{-1}$ cm$^{2}$ and ecf$_{HRC-I-2}$ = 7.7 x 10$^{10}$ cts erg$^{-1}$ cm$^{2}$, respectively. For the $\it{Swift}-$XRT data, the ecf has been computed to be 1.57 x 10$^{10}$ cts erg$^{-1}$ cm$^{2}$.

The spectroscopic data were then obtained by XMM$-$Newton from December 28, 2009 until February 02, 2010 in five distinct observations using the European Photon Imaging Camera (EPIC). EPIC$-$pn \citep{2001A&A...365L..18S}, MOS1 and MOS2 \citep{2001A&A...365L..27T} CCD detectors are mounted on the three X$-$ray telescopes on XMM$-$Newton. The exposure times as well as the raw count rates obtained with these detectors are presented in Table 1. The XMMSAS version 10.0.0 was used to filter the standard pipeline event files, to generate images, light curves, spectra and 
\begin{table*}[ht]

      \caption[]{The spectral parameters obtained from the model fitting from XMM-Newton}
         \label{Table:2}
\centering
      
         \begin{tabular}{llllcccrr}
            \hline
          \hline\\

 OBSID & N$_{H,M~31}^{a}$ & kT$^{b}$ & $\Gamma^{c}$ & $\chi^{2}$/d.o.f. & R$_{in}^{d}\sqrt{cos(i)}$ & L$_{PO}^{e}$ & L$_{BB}^{f}$  & L$_{Total}^{g}$  \\
& (10$^{20}$ cm$^{-2}$) & (keV) & & & (km) & (10$^{39}$ erg s$^{-1}$) & (10$^{39}$ erg s$^{-1}$) & (10$^{39}$ erg s$^{-1}$)\\
 
\hline
\hline \\
0600660201 &   5.1$~\pm~$1.1 & 1.070$~\pm~$0.010 & 2.59$~\pm~$0.16 & 1318.87/1165 & 52.90$~\pm~$0.01 & 0.33$~\pm~$0.07 & 0.94$~\pm~$0.07 & 2.16$~\pm~$0.07  \\	
0600660301 & 7.1$~\pm~$1.2 & 0.993$~\pm~$0.016 & 2.58$~\pm~$ 0.13 &  963.62/991 & 58.89$~\pm~$0.02 & 0.32$~\pm~$0.05 & 0.71$~\pm~$0.05 & 1.49$~\pm~$0.05  \\ 	
0600660401 & 9.1$~\pm~$1.2 & 0.913$~\pm~$0.013 & 2.81$~\pm~$ 0.14 &  1046.74/953  &  57.32$~\pm~$0.02 & 0.33$~\pm~$0.04 & 0.36$~\pm~$0.04 & 1.16$~\pm~$0.04  \\	         
0600660501 &   5.3$~\pm~$1.0 & 0.815$~\pm~$0.014 & 2.51$~\pm~$0.12  & 962.61/938 & 61.57$~\pm~$0.02 & 0.23$~\pm~$0.03 & 0.27$~\pm~$0.03 & 0.71$~\pm~$0.03  \\	
0600660601 &   6.5$~\pm~$1.6 & 0.769$~\pm~$ 0.013 & 2.70$~\pm~$ 0.13 & 830.66/821 &  64.18$~\pm~$0.03 & 0.16$~\pm~$0.03 & 0.25$~\pm~$0.03 & 0.65$~\pm~$0.03  \\
\hline
\hline
 \end{tabular}    
\tablefoot{  
\tablefoottext{a}{External absorption column density.}
\tablefoottext{b}{Temperature of inner$-$disc from DISKBB.}
\tablefoottext{c}{Photon index from powerlaw model.}
\tablefoottext{d}{R$_{in}$ is inner radius of the multicolored blackbody disk, and $\it{i}$ is the inclination angle.}
\tablefoottext{e}{Absorbed luminosity contribution from the powerlaw model (hard component) in (0.2$-$10 keV) energy band. }
\tablefoottext{f}{Absorbed luminosity from the multicolored disk component within (0.2$-$10 keV) energy band.}
\tablefoottext{g}{Total unabsorbed luminosity from the source within (0.2$-$10 keV) energy band.}
 }
 \end{table*}
the appropriate detector response functions. The event files for all three cameras, pn, MOS1 and MOS2 were analyzed with the following parameter settings: For spectrum generation, we set $\texttt{“FLAG = 0”}$ to reject bad pixels, and  \texttt{PATTERN $\leq$ 4} for pn and \texttt{$\leq$ 12} for MOS1 and MOS2 event files were allowed to reduce noise in the data. Light curves were derived to investigate the variability in the source. Barycenter correction was performed using the \texttt{barycen} task in XMMSAS. The light curves were then corrected from various effects such as bad pixels, GTIs, vignetting using \texttt{epiclccorr}. 
\begin{figure}[h!]
  \centering
   \includegraphics[angle=90,scale=0.40]{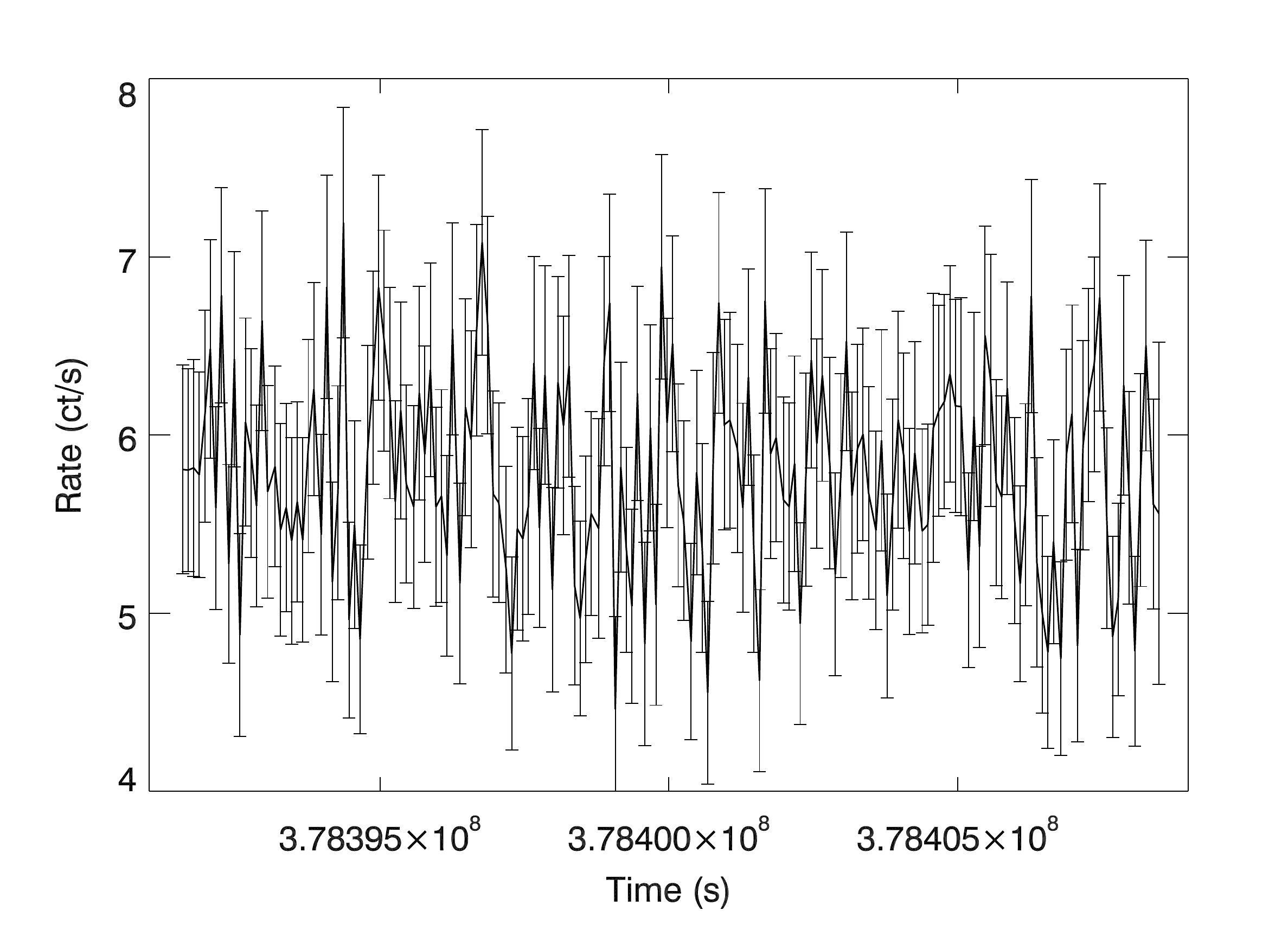}     
\caption{Light curve from XMM$-$Newton EPIC$-$pn for observation 0600220201 with time binning of 100 s in 0.2$-$ 10 keV energy band.}
               
  \end{figure}
As an example Fig. 2 displays the observed count rate for XMM$-$Newton EPIC-pn for the first observation, binned in 100 s intervals. The power density spectra were generated using the task  \texttt{powspec} in \texttt{FTOOLS} using the Fast Fourier transform algorithm (FFT). The normalization was chosen such that the white noise level expected from the data errors corresponds to a power of 2. %Fig. 9 shows the resulting power density spectra for the same observational bin shown in Fig. 8. 
For spectrum extraction, the XMMSAS tool \texttt{epatplot} was used to check for pile$-$up for appropriate source extraction region. The inner PSF part of these regions were excluded to avoid pile$-$up. Response files were generated by using SAS tools \texttt{rmfgen} and \texttt{arfgen}. The spectral binning was constructed to attain 20 counts per bin to assure uniform statistics across the energy range. Bad pixels were ignored and energy channels from 0.2 $-$ 10.0 keV were considered for spectral fitting using \texttt{XSPEC} version 12.6.0.

\section{Results}
 Our initial spectral fitting for XMM$-$Newton data used a broken powerlaw model (\texttt{BKNPOWER} in XSPEC), but the $\chi^{2}$/d.o.f. = 2953.96/1174 clearly indicated that this is not a satisfactory model. The luminosity derived from this simple fit indicated the fact that this source is a member of the ULX class. For these objects, an alternative, often better fitting model is a combination of a non$-$thermal (power law) and thermal component \citep{2009MNRAS.397.1836G}.  To explore this possibility we applied a model that combines a non$-$thermal powerlaw component (\texttt{PO}) with a multicolored blackbody component (\texttt{DISKBB}). The thermal component is associated with an accretion disk around a central black hole as developed by \citet{1984PASJ...36..741M} and \citet{1986ApJ...308..635M}. This model gave a significantly improved fit, indicated by $\chi^{2}$ / d.o.f. values in Table 2.

In addition to the intrinsic two component model, the source flux must be corrected for extinction by foreground gas in the Milky Way and the local gas along the line of sight through M~31. Extinction was modeled  using the Tuebingen$-$Boulder ISM absorption model (\texttt{TBABS}) \citep{2000ApJ...542..914W}. The hydrogen column density ascribed to the Milky Way was fixed to 5.32 x 10$^{20}$ cm$^{-2}$  \citep{1990ARA&A..28..215D}, whereas the column density for the host galaxy was included as a free  parameter. However, the derived value of the column density in M~31 must thus be considered with caution because we do not have information about the actual metallicity along the line of sight through M~31 and simply assume the Milky Way template expressed in terms of a standard cross$-$section per hydrogen atom.
% Combining all these factors, the spectral model follows from the following equation:

%$$f_{E}\, = e^{-\tau_{M31}(E)} \,e^{-\tau_{MW}(E)}} \,\sum\limits_{n} \,\phi_{n}(E,T_{in},R_{in},\alpha)\:$$ 
%\hspace{52 mm} photons cm$^{-2} s^{-1} keV^{-1}$ 
%\\

%where f$_{E}$ is the absorption corrected X$-$ray photon number flux , $\tau_{M31}$ and $\tau_{MW}$ are the opacities for the host galaxy and the Milky Way, respectively and the summation is over the model components used ($\phi_{n}$), which in turn are dependent on parameters such as $\alpha$ the photon index; $T_{in}$, the temperature at the inner radius of a multicolored blackbody disc. 
To illustrate this fitting procedure, Fig. 3 shows the XMM$-$Newton EPIC data for first of our five observational epochs. The decomposition of the X$-$ray spectrum into the model components in physical units is shown in Fig. 4.

\begin{figure}[h!]
   \centering
   \includegraphics[angle=270,scale=0.35]{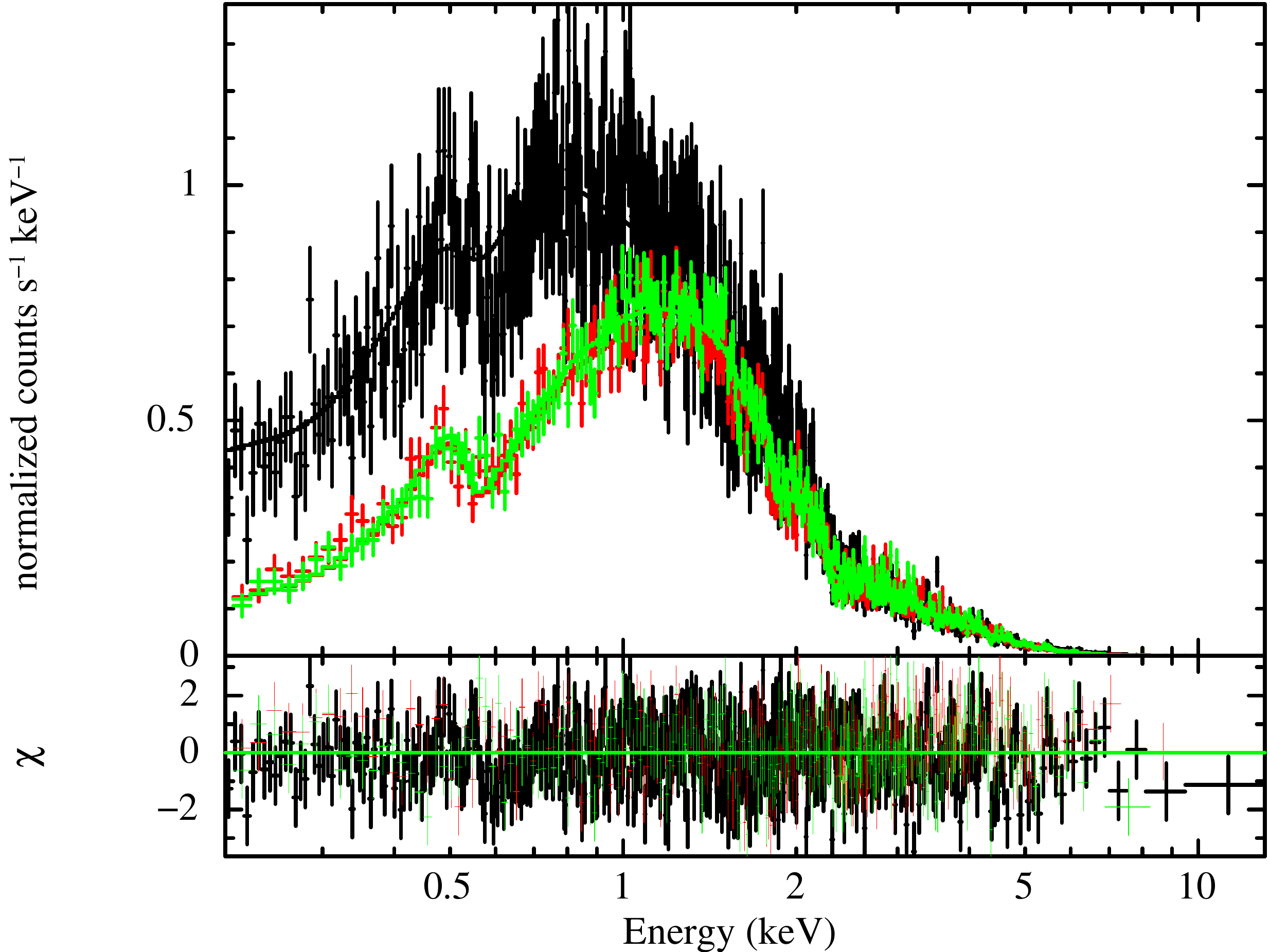}    
  \caption{Combined powerlaw and multicolored blackbody fit to the joint set of data from EPIC$-$pn (black), MOS1 (green), MOS2 (red)  with \ensuremath\chi$^2$/ d.o.f. = 1318.87/1165 for the observation 0600660201 with XMM$-$Newton. The model parameters obtained from this spectral fit are presented in Table 2.}
               
   \end{figure}
\begin{figure}[h!]
   \centering
   \includegraphics[angle=270,scale=0.35]{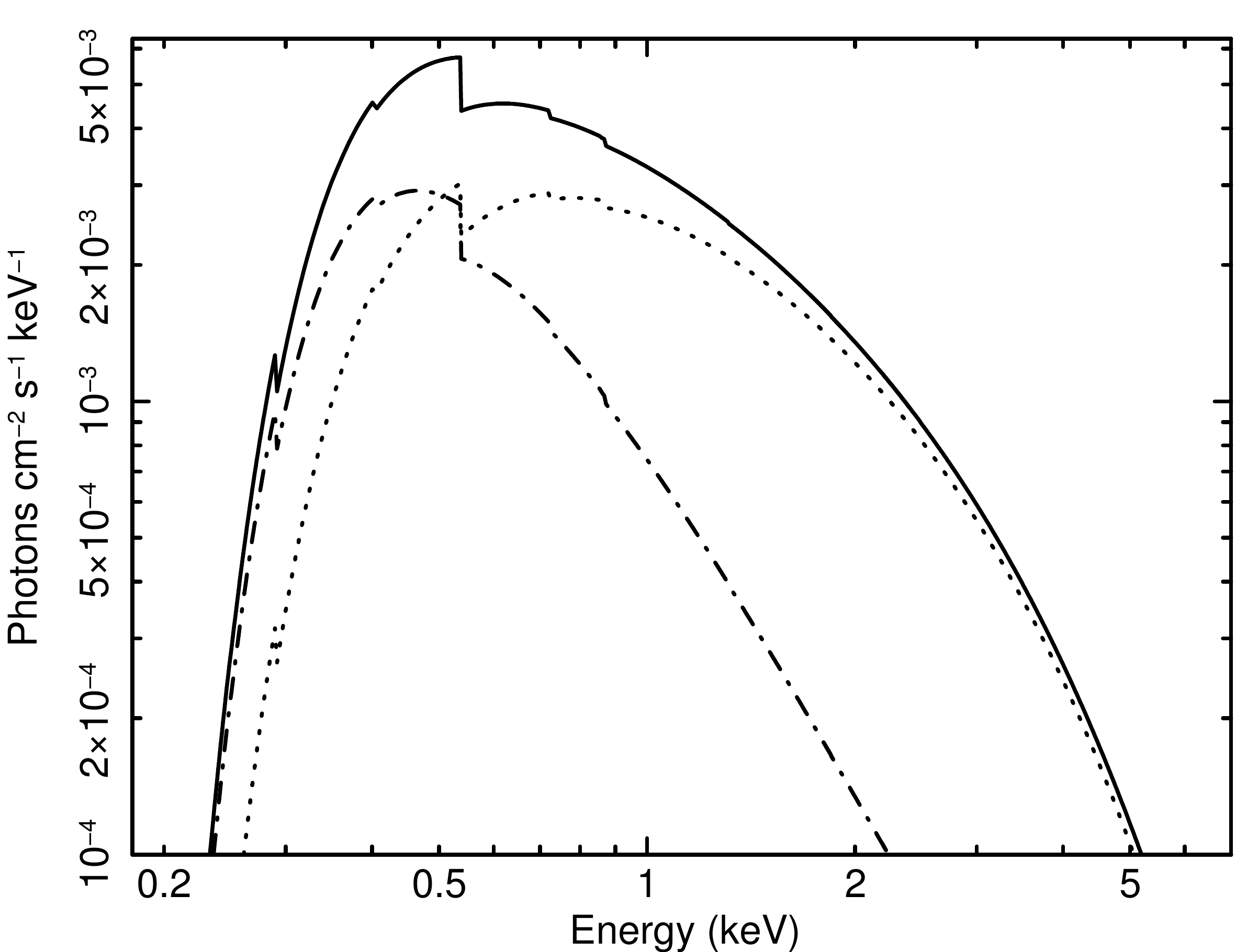}    
  \caption{The theoretical model underlying the data fit (solid line) in Fig. 3 is shown in decomposition between the thermal (dash-dotted line)and the non$-$thermal (dotted line) component.}
               
   \end{figure} 
To convert fluxes to luminosities, we assume a source distance of 780~kpc \citep{1998AJ....115.1916H, 1998ApJ...503L.131S} and isotropic emission. The resulting luminosities during the XMM$-$Newton observations are about 10$^{39}$ erg s$^{-1}$, which places this source in the mid range of known ULXs \citep{2009MNRAS.397.1836G}. Table 2  summarizes the derived luminosities (from XMM$-$Newton only) as well as the effective inner disk radius.  As a function of time, both the  non$-$thermal (L$_{PO}$) and thermal (L$_{BB}$) component decrease, but the latter decreases more rapidly (see Fig. 5). Table 1 lists the calculated luminosities derived from count rates associated with observations of $\it{Chandra}$ and $\it{Swift}-$XRT. The total unabsorbed luminosity declined from about 3.7 x 10$^{39}$ to 0.5 x 10$^{39}$ erg s$^{-1}$ during the period of observation. We note that this source was initially more luminous than the total X$-$ray luminosity of M~31 in the (0.1$-$2.4) keV band sampled by ROSAT \citep{1997A&A...317..328S}. Futhermore, this source was atleast one order of magnitude brighter than any of the 45 X$-$ray transients detected in M~31 by $\it{Chandra}$ and XMM$-$Newton from October, 1999 to August, 2002 \citep{2006ApJ...643..356W}. 

%\begin{table}[h!]
%\caption[]{The luminosity values obtained from $\it{Swift}-$XRT and $\it{Chandra}$~HRC-I} 
%\label{Table:3}
%\centering
      
 %        \begin{tabular}{lll}
  %          \hline
   %       \hline\\
%Telescope/Instrument & OBSID & L$_{unabs}^{a}$ & \\
%\hline
%\hline
%\\
%$\it{Chandra}$~HRC-I & 10885 & 0.002$^{b}$  \\
%$\it{Chandra}$~HRC-I & 10886 \\
%$\it{Swift}-$XRT & 31518013 & 3.04$~\pm~$0.20\\
%$\it{Swift}-$XRT & 35336016 & 3.31$~\pm~$0.10 \\
%$\it{Swift}-$XRT & 35336017 & 2.86$~\pm~$0.10\\
%$\it{Swift}-$XRT & 35336018 & 2.76$~\pm~$0.10\\
%$\it{Swift}-$XRT & 35336019 & 2.58$~\pm~$0.10\\
%$\it{Swift}-$XRT & 35336020 & 2.67$~\pm~$0.10 \\ 
%$\it{Chandra}$~HRC-I & 11808 & 0.514$~\pm~$0.08\\
%$\it{Chandra}$~HRC-I & 11809 & 0.460$~\pm~$0.18 \\ 
%\hline
%\hline
% \end{tabular}    
%\tablefoot{ The order is determined by the date of observation. 
%\tablefoottext{a}{Total unabsorbed luminosity from the source in units of 10%$^{39}$ erg s$^{-1}$.}
%\tablefoottext{b}{3$\sigma$ upper limit.}
% }
 %\end{table}  

\begin{figure}[h!]
  \centering
   \includegraphics[angle=0,scale=0.55]{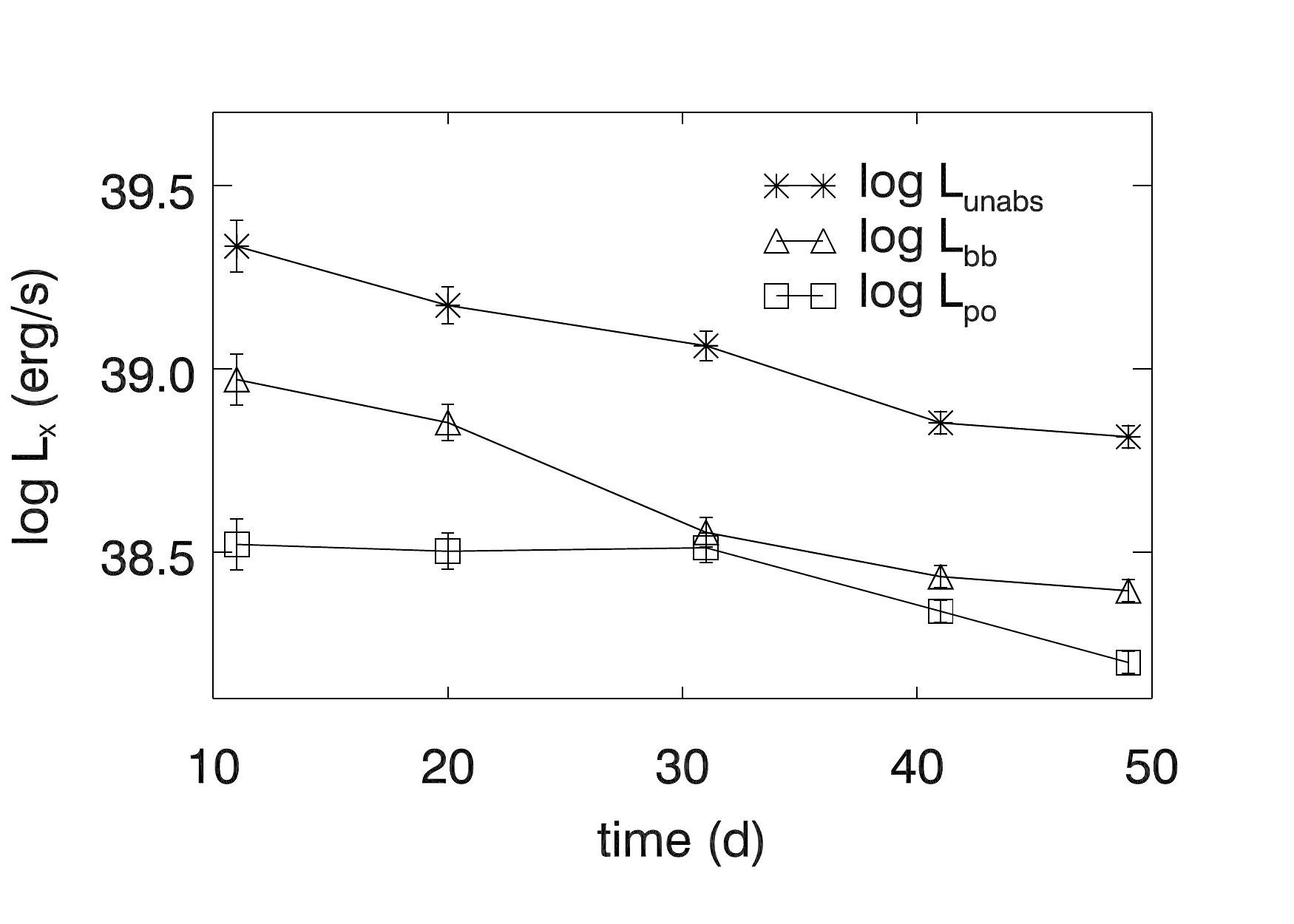}     
\caption{ Temporal variation of the total unabsorbed luminosity (L$_{unabs}$) as per observations of XMM$-$Newton. The decomposition into the powerlaw (L$_{po}$) and blackbody component (L$_{bb}$) reveals a distinct evolutionary behavior of these two components. The time zero corresponds to $\it{Chandra}$ HRC-I observation ID 10886 (see Table 1).}
               
   \end{figure}

From the X$-$ray spectrum fits of the first XMM$-$Newton observation (when the source was nearly brightest and we had the best quality data), we extract an estimate of the mass of the black hole following the formalism given by \citet{2000ApJ...535..632M}. Under the assumption of a geometrically flat and optically thick accretion disk, these authors derived the following relation:
$$ M =\frac{R_{in}}{8.86 \,\alpha} \,M_{\sun}\:,$$

where R$_{in}$ is the inner radius of the accretion disk,  measured in km and $\alpha$ is a dimensionless parameter that relates the inner radius to the Schwarzschild radius via R$_{in}$=3$\alpha$ R$_{S}$.  

For the bolometric luminosity emitted by the accretion disk, \citet{2000ApJ...535..632M} find (their eqt. 9):
$$ L_{bol} = 7.2 x 10^{38} \left(\frac{\xi}{0.41}\right)^{-2}\left(\frac{\kappa}{1.7}\right)^{-4}\alpha^2\left(\frac{M}{10 M_{\sun}}\right)^2\left(\frac{T_{in}}{keV}\right)^4 \:$$ 
\hspace{75 mm} erg s$^{-1} $
\\

where  T$_{in}$ is the temperature characterizing the inner accretion disk. The remaining parameters specific to their model are here chosen to take their standard values as indicated by the normalization. The spectral fit yields the products of R$_{in}$ and the square root of the cosine of the inclination angle of the disk, as well as temperature, T$_{in}$, and the bolometric luminosity (which we assume to be unabsorbed L$_{bb}$ component (= 1.6 x 10$^{39}$ erg s$^{-1}$) integrated from 0.2$-$10.0 keV). From the two equations above, we can therefore determine the mass and inclination angle, although significant uncertainties in both quantities are associated with the model parameters, $\alpha$, $\xi$ and $\kappa$ fixed to the values 1, 0.41 and 1.7, respectively \citep[see][]{2000ApJ...535..632M}. The $\alpha$ parameter has been defined above and $\xi$ is a correction factor, reflecting the fact that T$_{in}$ occurs at a radius somewhat larger than R$_{in}$. The specific value adopted here was motivated by \citet{1998PASJ...50..667K}. The final paramter, $\kappa$ represents the ratio of color temperature to effective temperature and the value adopted here has been taken from \citet{1995ApJ...445..780S}. With these caveats in mind, we find a mass of 13  M$_{\sun}$ with a statistical uncertainity of about 4 \% . This implies an inner radius of R$_{in}$ = 124 km. From the fitting parameter R$_{in}\sqrt{cos(i)}$ (see Table 2), one then infers an inclination angle of $\it{i}$ $\sim$ 80$^{\circ}$. The assumption of a non-rotating Schwarzschild black hole ($\alpha=1$) can lead to a significantly underestimated black hole mass as the extreme case of a maximally rotating Kerr black hole ($\alpha = {1}/{6}$) allows for a more massive central object approaching the realm of IMBH. Note, that this increase with black hole spin relies on the assumption that the inner edge of the accretion disk can be identified with the location of the innermost stable circular orbit (ISCO). However, determining the inner edge of a black hole accretion disk is a far more complex issue, e.g. \citet{2010A&A...521A..15A}. %However, the absence of a strongly broadened Fe K-$\alpha$ emission line in the spectra is an argument against a high spin, although the low signal-to-noise at this energy does not allow a firm conclusion.

As described in Sect. 2, we searched for an underlying periodicity in each of the observing runs over a range of frequencies limited by the temporal extent of the observation and the binning time of 100 s. The observed count rates (e.g. Fig. 2) display fluctuations within the range expected from counting statistics. None of the power density spectra reveal a significant feature.%, and all resemble the one shown in Fig. 9. 
\begin{figure}[H]
  \centering
   \includegraphics[angle=270,scale=0.30]{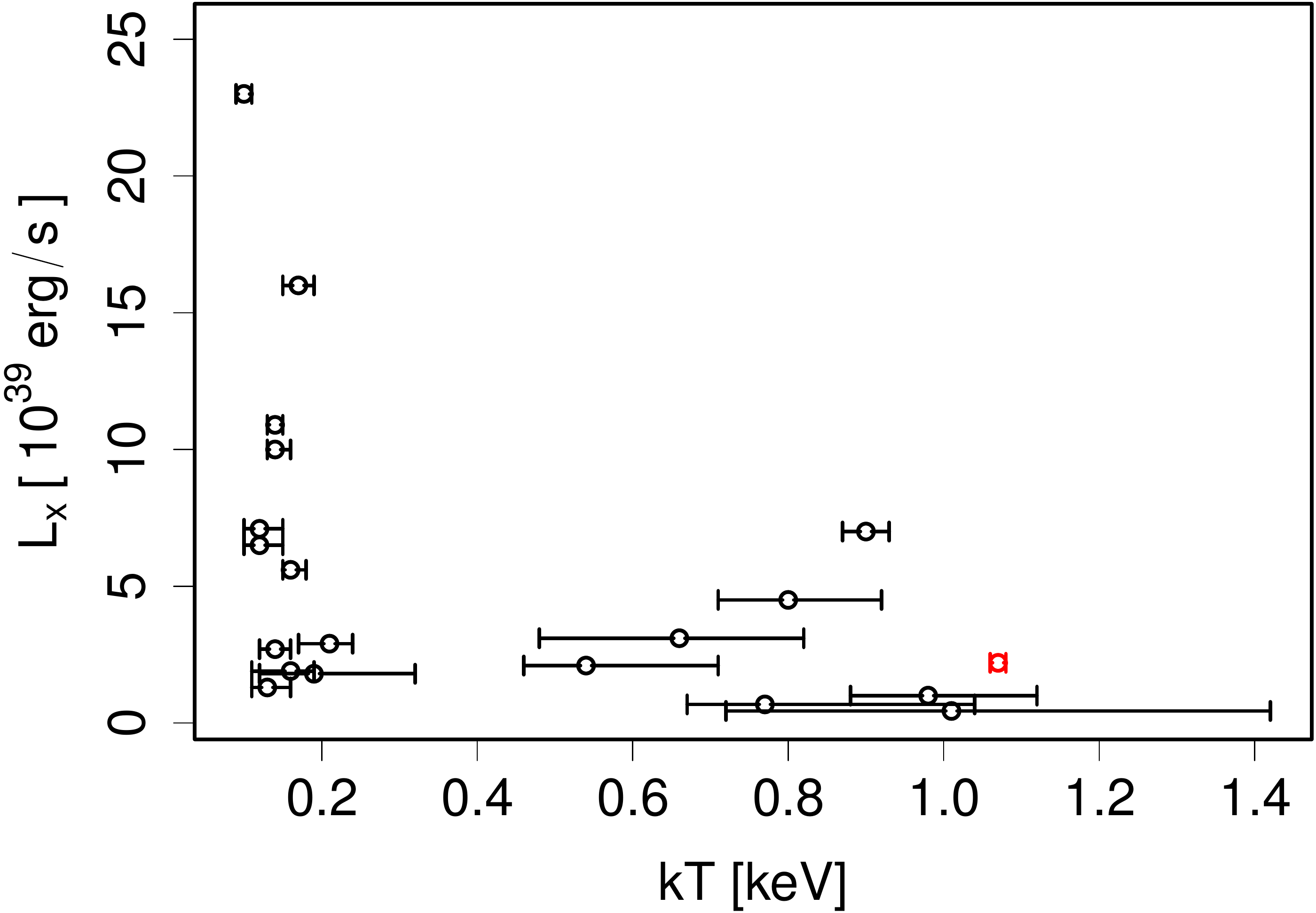}     

\caption{Temperature vs Luminosity for 19 selected ULXs from \citet{2006ApJ...649..730W} along with the candidate ULX (red) in M~31. The M~31 source is a member of the "high temperature" subgroup.}
               
   \end{figure}

\section{Discussion and conclusions}

The intrinsic luminosity of the source is consistent with those found for other ULXs in nearby galaxies. The presence of a thermal component at kT $\sim$ 1 keV in conjunction with a non$-$thermal high$-$energy tail, is also consistent with spectral properties of other ULXs. The temperature and photon index of M~31 source fall in the center of observed distributions for ULXs \citep{2006ApJ...649..730W}, lending additional support to a ULX nature.

We compared various properties of this ULX with a selected sample from the XMM$-$Newton archival study of ULXs by \citet{2006ApJ...649..730W} in 32 nearby galaxies with distance $<$ 8~Mpc and unabsorbed luminosities, $L_{x} > 10^{38} $ erg s$^{-1}$ in the 0.3 $-$ 10 keV energy range. Motivated by the observed properties of black hole binaries in the Milky Way, these authors classified their sample into two subsets, "low state" and "high state", based on their respective spectral properties. Low state sources were characterized by a single  powerlaw X$-$ray spectrum, while high state sources required an additional multicolored blackbody with powerlaw. Since the spectra described in Sect. 3 clearly exhibit the best fit when a powerlaw is combined with multicolored blackbody, this source should be classified as "high state" ULX. Therefore we limit our  comparison to 19 of high state ULXs in their sample. This sample is slightly reduced because we selected the brightest epoch out of multiple XMM$-$Newton observations, and those were not always available in \citet{2006ApJ...649..730W}.

Figure 6 displays the resulting set of "high state" ULXs in the L$-$T plane. As \citet{2006ApJ...649..730W} pointed out, the temperature distribution is possibly bimodal. The ULX in M~31 belongs to the high-T group, which is commonly believed to represent the class of stellar mass black holes. The low$-$T group is characterized by a larger range of X$-$ray luminosities and has been tentatively attributed to the hypothetical class of IMBHs \citep{2006ApJ...649..730W}. Although the sample size is still very small, it appears that a physical gap between these two groups exists. If that is the case, an explanation may be related to different formation scenarios for these groups. As \citet{2006ApJ...649..730W} suggested, one possibility is associated with the unique features of pair instability supernovae (PISNe) in the early universe (Population III origin, \citealt{2003ApJ...591..288H}). The mass of $\sim$ 13 M$_{\sun}$ derived in Sect. 3  suggests that we are dealing with a normal stellar remnant, one that could have been formed later in the universe, provided $\alpha$ is not significantly less than unity due to rotation. 

\begin{figure}[H]
  \centering
   \includegraphics[angle=270,scale=0.30]{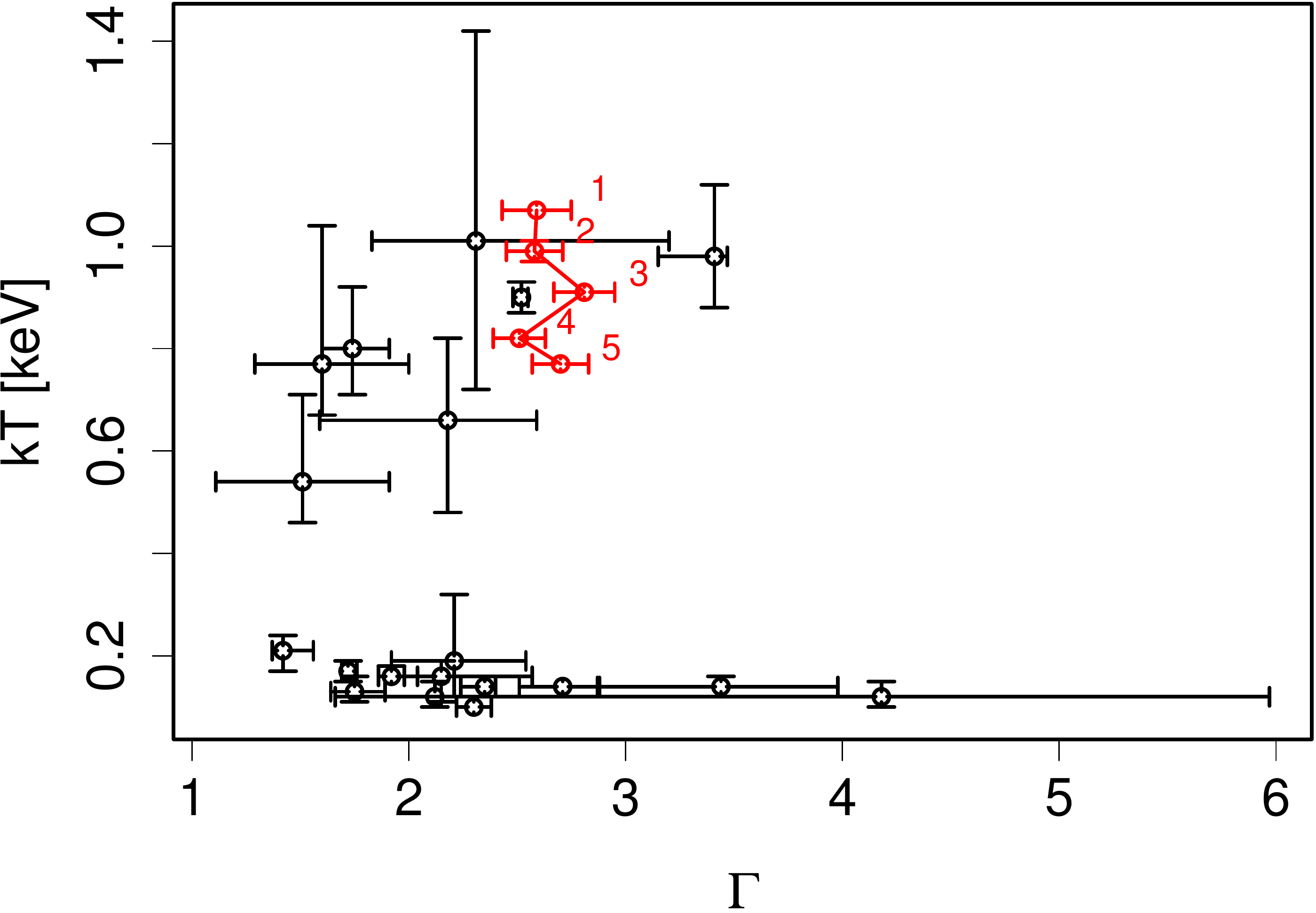}     

\caption{For the same sample shown in Fig. 6, a comparison of photon index vs temperature reveals that the M~31 ULX cools significantly, but remains within the range covered by the "high temperature" group. The labels 1 through 5 correspond to the chronological entry of XMM$-$Newton observations as listed in Table 1.}
               
   \end{figure}

\begin{figure}[h!]
  \centering
   \includegraphics[angle=180,scale=0.35]{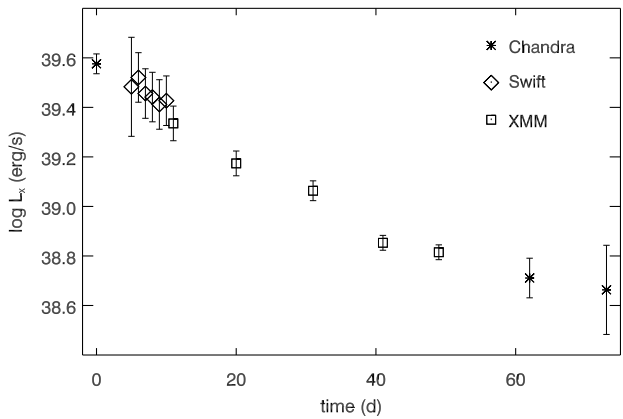}     
\caption{Variation of the total unabsorbed luminosity (L$_{unabs}$) of ULX in M~31 as observed by $\it{Swift}-$XRT, $\it{Chandra}$~HRC-I and XMM$-$Newton resembling an exponential decay with a time scale of 32 days. The time zero corresponds to $\it{Chandra~HRC-I}$ observation ID 10886 (see table 1).}
               
  \end{figure}

Figure 7 shows the "high state" ULXs distribution in the T$-\Gamma$ plane. The low-T group shows a large range in photon index, resembling the large range in luminosity, however we note that the uncertainties in this parameter are very large. In contrast to the previous figure, here we show the temporal evolution of the M~31 ULX, while for other ULXs, no evolution has been shown. Over a time span of forty days of observations with XMM$-$Newton, the spectral evolution clearly exhibits cooling, but almost no change in the shape of the power-law component. Referring back to Fig. 5, the drop in the temperature is also accompanied by a significant change in the luminosity of this component. The power-law component, on the other hand, exhibits a declining luminosity as well, but not as rapid as the thermal one. The overall evolution of this transient is thus a combination of a strongly varying thermal component and a significantly less varying non$-$thermal part.

Combining all data obtained with \emph{Chandra}, \emph{Swift} and XMM$-$Newton indicates that the X$-$ray luminosity of this source follows closely an exponential decline (see Fig. 8) with a time constant of 32 days. A continuation of this trend is evident from data taken 150 days after the outburst when the source was detected at $\sim$ 10$^{38}$ erg s$^{-1}$ \citep{2011ApJ...734...79B}. This observed exponential decline of the luminosity is consistent with those determined for the class of galactic X$-$ray novae \citep{1997ApJ...491..312C}, which show FRED (Fast Rise Exponential Decay) like light curves, though the FR part in our case was missed. However, the light curves of this source class often exhibit more complex behaviour than a simple exponential decay.

In summary, the $\it{Chandra}$ discovery of CXOM31~J004253.1+411422 has established it as the first member of the class of ULXs in M~31. Follow up observations with $\it{Swift}$ and XMM$-$Newton revealed an exponential decline, reminiscent of the late time evolution of Galactic X$-$ray novae. Spectral analysis of the XMM$-$Newton data suggests that this source was in a "high state" at the time of observation, and that the underlying source is likely a stellar mass black hole accreting near the Eddington limit.

%\begin{figure}[h!]
%  \centering
%   \includegraphics[angle=270,scale=0.40]{pds.pdf}     
%\caption{PDS from XMM$-$Newton showing no characteristic peak over the given frequency range.}
               
%  \end{figure}

\begin{acknowledgements}
 The XMM project is supported by the Bundesministerium f\"{u}r Bildung und Forschung / Deutsches Zentrum f\"{u}r Luft- und Raumfahrt (BMBF/DLR) and the Max-Planck Society. We thank the $\it{Swift}$ team for their help with scheduling the TOO observations. M. Henze acknowledges support from the BMWI/DLR, FKZ 50 OR 1010.

\end{acknowledgements}

\bibliography{ulx-paper}
\end{document}